\documentstyle[12pt,aasms4]{article}
\begin{document}
 
\title{Results from GROCSE: A Real-time Search for Gamma Ray Burst
Optical Counterparts}

\author{Brian Lee\altaffilmark{1}, 
Carl Akerlof\altaffilmark{1}, 
David Band\altaffilmark{2}, 
Scott Barthelmy\altaffilmark{3,4}, 
Paul Butterworth\altaffilmark{3,5},
Thomas Cline\altaffilmark{3}, 
Donald Ferguson\altaffilmark{6}, 
Neil Gehrels\altaffilmark{3}, 
Kevin Hurley\altaffilmark{7}.}

\altaffiltext{1}{University of Michigan, Ann Arbor, MI 48109}
\altaffiltext{2}{CASS, University of California San Diego, La Jolla, CA 92093}
\altaffiltext{3}{NASA Goddard Space Flight Center, Greenbelt, MD 20771}
\altaffiltext{4}{USRA, Columbia, MD 21044}
\altaffiltext{5}{HSTX, Lanham, MD 20706}
\altaffiltext{6}{EarthWatch Inc., Pleasanton, CA 94566}
\altaffiltext{7}{Space Sciences Laboratory, University of California,
Berkeley, CA 94720}

\received{1997 February 19}
\accepted{1997 March 27}
\bigskip
\centerline{To appear in {\it The Astrophysical Journal Letters}}

\begin{abstract} 

The Gamma-Ray Optical Counterpart Search Experiment (GROCSE) has
searched for contemporaneous optical counterparts to gamma ray bursts
(GRBs) using an automated rapidly slewing wide field of view optical
telescope at Lawrence Livermore National Laboratory.  The telescope
was triggered in real time by the Burst And Transient Source
Experiment (BATSE) data telemetry stream as processed and distributed
by the BATSE COordinates DIstribution NEtwork (BACODINE).  GROCSE
recorded sky images for 28 GRB triggers between January 1994 and June
1996.  The analysis of the 12 best events is presented here, half of
which were recorded during detectable gamma ray emission.  No optical
counterparts have been detected to limiting magnitudes $m_V \leq 8.5$
despite near complete coverage of burst error boxes.

\end{abstract}


\section{Introduction}

  
Nearly three decades after their discovery, the phenomenon of gamma
ray bursts is still an outstanding mystery.
Results from the Burst and Transient Source
Experiment (BATSE) (Fishman et al. 1989; Meegan et al. 1996) aboard
the Compton Gamma Ray Observatory (CGRO) have shown that the sources
are isotropically distributed on the sky, limited in radial extent,
and from quite compact sources which are most likely outside the disk
of our galaxy.  No theory has yet been proposed which satisfactorily
describes all the observed characteristics.
Optical counterpart searches could reveal two additional kinds of
information inaccessible to BATSE that might resolve the mystery.
First, the BATSE error boxes are several degrees across, including
countless possible astrophysical objects.  Positive detections of
transient optical counterparts would pinpoint source coordinates to
arc-second accuracy and possibly identify quiescent sources.  Second,
gamma ray bursts have only been observed in the gamma- and X-ray
energy bands.  Additional spectral information can significantly
constrain models of the production mechanism.

No optical counterparts of $m_V~\le$ 24 (Schaefer 1994) have appeared
in follow-up observations made from hours to days after bursts.
Archival searches for repeating counterparts have yielded some
potential non-contemporaneous candidates (Schaefer 1991; Hudec \&
Sold\'{a}n 1994) but the lack of temporal correspondence has left
these candidates in doubt.  Simultaneous observations have been
attempted by groups such as the Explosive Transient Camera (ETC)
(Krimm et al. 1996) and the Ond\v{r}ejov network (Hudec et al. 1984).
But sufficiently deep wide field of view stationary instruments are
prohibitively expensive, and even with a steradian field of view,
would only observe one or two events per year.  Fortunately, following
the tape failure onboard CGRO, real time telemetry was required, and
the BATSE COordinates DIstribution NEtwork (BACODINE) (Barthelmy et
al. 1994) was created in 1993 to distribute real time BATSE trigger
information.  Thus the GROCSE collaboration (Akerlof et al. 1994,
1995; Lee et al. 1996; Park et al. 1996) commenced an ambitious
contemporaneous optical counterpart search program.  With its wide
field of view and rapid slewing abilities, GROCSE was able to observe
the entire error box of several bursts a year within seconds of their
detection by BATSE, and in many cases while BATSE was still recording
gamma ray emission.  No counterparts were detected.


\section{The GROCSE Instrument}

In the summer of 1993, we initiated the GROCSE collaboration to adapt
a camera array originally designed for the Strategic Defense
Initiative program to the task of detecting GRBs.  This
Wide-Field-Of-View (WFOV) system was designed to be a high frame rate,
multiple target tracking instrument (Park et al. 1989, 1990), and thus
had requirements rather different from those of a normal CCD camera.
The WFOV system consisted of a single 60$^\circ$ wide-field-of-view
lens, 23 image intensified CCD imaging systems, and associated readout
electronics.  The $f/2.8$ lens had an effective aperture of 89 mm and
a focal length of 250 mm.  Twenty-three reducing fiber optic bundles
transported light from the spherical focal plane to image
intensifiers.  The output of each intensifier was transmitted through
a second reducing fiber bundle to a 384$\times$576 CCD covering a
7.5$\times$11.5$^\circ$ field of view.  The 23 cameras covered 75\% of
the lens' field of view for an effective solid angle of 0.621
steradians.  The exposure duration was set to 0.5 second shuttered by
the image intensifiers.  The quantum efficiency was constrained by the
limited acceptance of the fiber optic reducing bundles and the
photoelectric conversion in the image intensifiers, which combined
with the local bright sky background, lead to a detection threshold of
$m_V = 8.5$.  The entire assembly was mounted on a Contraves
computer-controlled inertial guidance test system.

The analog CCD outputs were serially multiplexed into a single
Datacube image processing system under the control of a Sun 4/330 host
computer.  To compensate for saturation and variable brightness of the
night sky, the intensifier gain was dynamically adjusted, camera by
camera, to achieve the best performance. The spectral response of the
GROCSE image intensifiers extended from 400 to 900 nm with peak
sensitivity around 650 nm according to data supplied by the
manufacturer. This was experimentally verified by comparing the GROCSE
response for stars with widely different surface temperatures.

GROCSE operated in two modes: sky patrol and burst.  It spent most of
its time in sky patrol mode using all 23 cameras and seven camera
array positions to image the fraction of the sky with elevation angle
greater than 30$^\circ$ every half hour.  (Adjacent buildings obscured
most of the sky below 30$^\circ$.) GROCSE skipped any position within
30$^\circ$ of the moon in order to protect the image intensifiers from
damage.
Burst mode occurred whenever GROCSE received a burst trigger via
BACODINE.  The GROCSE host computer maintained a link to BACODINE 24
hours a day, trading packets once per minute.  BACODINE is capable of
distributing BATSE burst trigger information and approximate
coordinates within 6 seconds after the BATSE trigger although for some
slow rise triggers the delay exceeded 20 seconds.  If GROCSE received
a trigger with coordinates safely distant from the moon, any current
operation was interrupted and the system started burst mode.  The
camera array would slew to the approximate burst position in less than
10 seconds and take images as rapidly as possible for 20 minutes.  If
the sky conditions had changed significantly from previous exposures,
the auto-gain code would require a few additional seconds of
processing time before the first image was recorded.  The median
response time for BACODINE triggers was 16 seconds from the initial
detection onboard the CGRO to the first GROCSE image.  After recording
images for 20 minutes, the system returned to sky patrol mode.  This
protocol insured a substantial number of images both before and after
each burst.

\section{Data Analysis}

Analysis of the GROCSE data required several steps.  First, the
relative location of each camera with respect to the central one was
determined by comparison of star images to corresponding elements of
the SAO catalog.  Next, ``warp'' maps were generated to correct the
2-D distortions introduced by the intensifiers and fiber optic
reducing bundles.  These warps, unique to each camera, were typically
on the order of 6 arc-minutes but could be as large as 20 arc-minutes.
By examining several images and measuring the residual between the
apparent and real positions of stars, our code created a 23$\times$36
warp map for the entire camera field with which we could correct a
star's position to within 3 arc-minutes.  These first two calibration
steps had to be done only once for the entire analysis.  The only
remaining degree of freedom was the position of the entire camera
array for each event.

Once the star images were accurately mapped to celestial coordinates,
the BATSE error box (Meegan et al. 1995) was examined for transients
using the BATSE 3-$\sigma$ statistical (three times the quoted
1-$\sigma$ error) plus 1.6$^\circ$ average systematic errors added in
quadrature.  When IPN (Interplanetary Network) arcs (Hurley et
al. 1994) were available, we further limited the search region to the
overlap between this arc and the BATSE error box.  Often, the error
box covered more than one camera.

To search for transients in the image, our code first performed a
``pseudo field flattening'' to correct for variations of intensifier
sensitivity across the image.  (Since intensifier response varied with
gain which was automatically adjusted by the online code to optimize
peak image quality, we could not use true flat field images.)  All
pixels al least 5-$\sigma$ above the surrounding background were then
identified, and the centroids of all connected clusters of pixels were
computed.

Using the camera alignment parameters and warp maps previously
determined, the coordinate transform code automatically found the
absolute RA and DEC coordinates of each pixel cluster centroid in all
of the images.  Any cluster would be considered a counterpart
candidate if it could not be identified with an SAO cataloged star and
was present in at least two consecutive frames, ruling out satellite
glints or cosmic ray hits.

Since GROCSE did not track the sky, stars moved from one image to the
next while pixel defects did not.  Thus we could compare these
remaining clusters to similar lists for sequential images and identify
them as belonging to one of three classes: star-like objects, bad
pixels, and single-frame noise.

Star-like objects were those which retained the same RA and DEC in
multiple images, and thus were likely to be real celestial bodies.
These were identified by comparison to other star and object catalogs
as well as to sky patrol images from earlier and later in the same
night and other nights.  A few were galaxies or variable stars but
most were simply normal stars near threshold but not included in the
SAO catalog.  Any such object we could not identify would have been a
potential counterpart.  However, all such objects eventually proved to
be consistent with known celestial bodies.  Once we had identified a
repeating pixel cluster, we entered it into our own catalog, and
compared all subsequent images to both the SAO catalog and our own.

Bad pixels were pixels which always had much higher counts than any
surrounding pixels, a common problem in the GROCSE system.  Unlike hot
pixels in a normal CCD camera system, the counts of these bad pixels
could still vary widely from frame to frame.  Once identified, these
locations (at most a few hundred per camera) were appropriately
cataloged.

Single-frame noise consisted of any cluster of pixels which appeared
above threshold in only one frame, matching no other cluster in pixel
position or RA and DEC.  The majority of these clusters were ``bad
pixels'' as described above which were a few standard deviations above
all surrounding pixels in most frames, but more than 5-$\sigma$ above
the background for only one frame, and thus could be identified by the
code.  A few were dim stars visible in many frames but above
5-$\sigma$ for only one, and were identified by hand.  A very small
number were satellites, which we easily detected by their streaked
appearance and passage through predictable positions in several
cameras across our total field of view.  All single-frame objects were
identified leaving no possible candidates for an optical counterpart
for any of these events.

\section{Results}

GROCSE recorded images for BACODINE triggers corresponding to 28
actual bursts during its operation from January 1994 through June
1996.  BACODINE sent two kinds of burst triggers: ``type 1'' triggers,
which were received approximately 6 seconds after the BATSE detection;
and ``type 11'' triggers, which arrived some 10 minutes later.  Twelve
of the twenty-two ``type 1'' triggers were useful for later analysis;
the rest were marred by clouds or large BACODINE coordinate errors.
Prior to January 1995, GROCSE used only the 7 central cameras in burst
mode.  Since the final burst position was often outside of this field,
we switched to using all 23 cameras, greatly increasing our coverage
of many events but also increasing the readout time between
consecutive images.  Burst 940129 and all post-1994 bursts had on the
order of 25 seconds separating consecutive images in the same camera;
the other three 1994 bursts had consecutive images separated by an
average of only 5 seconds.
For the events analyzed, images were sequentially taken during and
after the burst for typical durations of 20 minutes.  (Only 10 minutes
of data are available for burst 940129.)  


Results and limits are listed in table I.  The $T_{90}$ values listed
are from the BATSE 3B catalog (Meegan et al. 1995) for the first four
bursts, and are rough duration estimates from the BATSE Collaboration
for later bursts.  ``First image'' is the time after the BATSE burst
trigger that the GROCSE camera overlapping the burst location was
recorded.  Coverage was determined primarily by modeling the BATSE
error box with a circular Gaussian probability surface.  When IPN arcs
were available, they were modeled with Gaussian errors and used to
restrict the error box further.  The 6 arc-minute square error boxes
used for matching to the SAO catalog effectively mask part of the
image, approximately 5\% in most cases, so this area must be included
in correcting estimates of the detection probability.

The failure to detect an optical transient provides us with an upper
limit on any possible optical emission at the time of the image.
First, we have listed approximate limiting V-band magnitudes.  A
comparison of the V-band filter to the GROCSE response function showed
that for most stellar types the calibrated GROCSE response would be
within 0.1 magnitudes of the listed SAO catalog $m_V$.  Stellar
classes B and K could vary by up to $\pm$0.3 magnitudes, as GROCSE is
relatively more sensitive to red stars than the V-band.  Independent
of color, observed magnitudes of a single star often varied by as much
as $\pm$0.5 from one frame to the next due to fluctuations in
intensifier sensitivity and discontinuities in optical collection
across the fiber optic bundles.  Thus any comparison to V-band
magnitudes is approximate at best.

The quoted limiting visual magnitude was determined by comparison to
SAO catalog stars.  The percent of detected SAO stars of a given
visual magnitude falls from near 100\% to near zero over slightly more
than a magnitude of brightness for most GROCSE images.  After binning
stars in 0.1 $m_V$ intervals, we set our limit at the dimmest
magnitude where at least 50\% of the SAO objects of the same $m_V$
were detected.  This corresponded to the detection of approximately
85$\pm$5\% of the SAO stars brighter than or equal to the listed
limit.  In most cases, some stars one or more visual magnitudes dimmer
than the quoted limit were visible in the image, but because the
system has much more noise than a normal CCD camera, such detections
are not reliable measures of sensitivity.  Under ideal conditions (a
moonless, clear night) this system could reliably identify stars as
dim as $m_V = 8.5$.  These limiting magnitudes have an error of
$\pm$0.5 $m_V$.

Our second optical emission limit is independent of the $m_V$ limit.
Since a spectral distribution must always be assumed, we adopt a
bremsstrahlung photon number density for the GRB spectrum,
$dN~\propto~dE/E$.  If the power law dependence of the spectral
distribution is drastically different, then either optical
counterparts would have been discovered long ago or will never be
detectable.  This distribution implies an integral flux
$N~=~\gamma~ln(E_b/E_a)$ where the bandpass is from $E_a$ to $E_b$,
and the constant $\gamma$ has units of $photons~cm^{-2}~s^{-1}$.  This
assumed distribution defines a robust measure insensitive to small
variations in the true spectral shape.  We convolve such a spectrum
with the GROCSE response curve to place an upper limit on
$\gamma_{opt}$ listed in table I.  This method is also applied to the
BATSE gamma-ray band to find $\gamma_{\gamma}$.

The GROCSE optical flux limit is determined directly from detector
response.  For each image we convolved the GROCSE response curve with
the black-body spectral distributions of a number of reference stars
with known spectral type and brightness.  The result was compared to
the actual detector response to estimate the optical flux
corresponding to a signal 5-$\sigma$ above the background noise.  This
determined the constant $\gamma_{opt}$ described above and the
limiting observable burst optical flux for the GROCSE waveband.  Note
that the uncertainty in this number is approximately a factor of 2 and
the real limit may be as much as 1 magnitude less than what is
reported.  When the error box was contained within several cameras,
limits for the camera with the highest detection probability are
given.

We list gamma ray fluences from 20 to 2000 keV integrated up to the
time of the first observation which in most cases was near the end of
the burst.  We used a number of methods to derive the best fluence
based on available data, as noted in table I.  The Band et al. four
parameter GRB model (1993) was used to fit BATSE High Energy
Resolution Burst (HERB) or Spectroscopy High Energy Resolution Burst
(SHERB) data (Fishman et al. 1989) when possible.  When these data
were unavailable, we used fluences from the BATSE 3B catalog (Meegan
1995) instead.  For the remaining bursts, we employed fits to
Spectroscopy Time-Tagged Event (STTE) data (Fishman et al. 1989).  The
STTE data can be accumulated over a variety of time ranges to form
spectra, but the data do not cover the entire burst and thus the
background subtraction for these events is less certain.  An estimate
of $\gamma_{\gamma}$ is then made using the fluence divided by the
time between the BATSE trigger and the GROCSE observation.

GROCSE imaged half of the dozen observed bursts while BATSE was still
recording gamma ray flux above background.  For four of these six
bursts, the GROCSE observation occurred late in the burst for the
cameras with high detection probabilities and difficulties with
background subtraction made determination of accurate simultaneous
gamma ray flux numbers impossible.  For the other two, 951124 and
951220, the flux was still high above background and, using the BATSE
HERB data type, we determined the simultaneous gamma ray flux.  This
is used below to find an instantaneous $\gamma_{\gamma}$ in addition
to the fluence over time to image figure used in the table.  For
951124 (Figure 1) the first camera imaged the burst at t+23 seconds,
and spatial coverage for that camera alone was 21\%.  Our
instantaneous fit for the gamma ray flux gives $\gamma_{\gamma} =
0.42~photons~cm^{-2}~s^{-1}$ at t+23 seconds, or a ratio
$\gamma_{opt}/\gamma_{\gamma} = 4700$.  The remainder of the error
box, with an additional 64\% coverage, was imaged by another camera at
t+29 seconds.  The instantaneous gamma ray fit at t+29 seconds gives
$\gamma_{\gamma} = 0.23~photons~cm^{-2}~s^{-1}$, or
$\gamma_{opt}/\gamma_{\gamma} = 9700$.  The second of these two
bursts, 951220, was imaged by a single camera with 95\% spatial
detection probability.  Our instantaneous gamma ray flux fit gives
$\gamma_{\gamma} = 0.21~photons~cm^{-2}~s^{-1}$ at the exposure time
of t+15 seconds, for a instantaneous ratio
$\gamma_{opt}/\gamma_{\gamma} = 15000$.

\section{Discussion}

The GROCSE contemporaneous optical counterpart emission limits are the
deepest yet established in this field, with the most extensive spatial
coverage.  While no single event can claim complete coverage, the
spatial probability of GROCSE having missed all 12 counterparts is
less than 1 in $10^{6}$.

ETC has reported similar visual magnitude limits for six bursts (Krimm
et al. 1996).  The ETC spatial detection probability is much less than
for GROCSE in most cases, but three of their simultaneous five second
exposures observed portions of the burst error box over the entire
period of gamma ray emission recorded by BATSE.  Their best published
of these limits is for burst 940305 for which they quote a limiting
$m_{V}$(1 sec flash) = 7.7 and an optical limit across their 400 to
750 nm bandpass of $2.8 \times 10^{-9} erg~cm^{-2}$.  This gives an
approximate ratio $\gamma_{opt}/\gamma_{\gamma} = 17500$.  Other ETC
events produce ratios similar to those reported here for GROCSE.

It is not entirely surprising that GROCSE and other experiments have
not yet detected optical counterparts.  Using the bremsstrahlung
assumption for the low energy GRB tail as discussed above, the very
brightest bursts would have optical counterparts of $m_V >$ 9, with a
typical burst producing an $m_V~\sim~14$ flash, and the brightest 10\%
of all bursts would have $m_V~\sim~10$.  Ford and Band (1996)
presented a more detailed prediction, fitting spectra to 54 bright
bursts using Band's four parameter model to extrapolate a flux in the
optical V-band.  Under the most optimistic assumptions, the results
predict, at best, that the brightest counterpart in their sample would
have been just beyond detection by GROCSE under even the best seeing
conditions and none of the bursts in the GROCSE sample could have been
detected.  GROCSE only observed approximately 2\% of the bursts
detected by BATSE, and only one of the 12 bursts was observed under
ideal conditions.  Thus the detection of a ``once a year'' burst
bright enough to produce a counterpart at the GROCSE detection limit
was unlikely in the two year span of these observations.

\section{Conclusions}

In over two years of operation, GROCSE has ruled out any possible
counterpart candidates for twelve bursts down to magnitudes as dim as
$m_V$ = 8.5 with near complete spatial coverage.  Further, GROCSE has
demonstrated the feasibility of simultaneous optical counterpart
searches using realtime BATSE data processed by BACODINE.  The GROCSE
limits are consistent with the results of Ford and Band (1996) which
predict observable optical counterparts no lower than $m_V~\sim$ 10.
Many GRB theories (M\'{e}sz\'{a}ros \& Rees 1996; Katz 1994) predict
optical counterparts of $m_V~\sim~9~-~14$ fading rapidly within a few
$\times$ 1 to $10^3$ seconds after the burst.  Faster and more
sensitive searches in this region are now commencing and should
provide results within a year.

We thank Gerald Fishman (NASA/MSFC), Chryssa Kouveliotou (USRA),
Charles Meegan (NASA/MSFC) and the entire BATSE team for their
assistance in this project as well as Elden Ables, Richard Bionta,
Linda Ott, Hye-Sook Park, and Eric Parker of Lawrence Livermore
National Laboratory.  Brian Lee is grateful for support by NASA grant
NGT-52805 provided by the Graduate Student Researchers Program.  Kevin
Hurley is grateful for support of the IPN under NASA Grant NAG 5-1560,
and for Ulysses operations under JPL Contract 958056.

\newpage

\figcaption{Coverage of the error box for GRB951124 in three GROCSE
camera images.  The BATSE error box (circle) is centered on right
ascension 73.3$^\circ$ declination 51.7$^\circ$ and has a 3-$\sigma$
statistical plus 1.6$^\circ$ average systematic error radius (added in
quadrature) of 3.8$^\circ$.  The IPN arc width is 0.18$^\circ$.  North
is approximately towards the top of the figure, and the bright star
visible in the lower left image is Capella.}

\newpage

\begin{deluxetable}{l c c c c c c c c c c c c}
\scriptsize
\tablecaption{GROCSE Observations}
\tablehead{
Name & Trigger & $T_{90}$ & First & IPN & Error & $m_V$ 
& $F_{opt}$\tablenotemark{a} & 20-2000 keV & 20-2000 keV
& $\gamma_{opt}$\tablenotemark{c} & $\gamma_{opt}/\gamma_{\gamma}$ \\
 & & or & image & & Box & limit & & Fluence\tablenotemark{b} 
& Fluence\tablenotemark{bf} & & \\
 & & duration & (sec) & & Coverage & & & $ergs~cm^{-2}$ & $photons~cm^{-2}$ 
& & \\
}
\startdata
940129 & 2793 &  7.0 & 35 & y & 80\% & 7.3 & 3200 
& $1.53\times 10^{-5}$\tablenotemark{e} & 16.1 & 4000 &  40000 \\
940623 & 3040 & 26.0 & 24\tablenotemark{d} & n & 77\% & 7.3 & 4100 
& $3.28\times 10^{-6}$\tablenotemark{f} & 30.9 & 5000 &  18000 \\
940828 & 3141 &  2.3 & 21 & n & 65\% & 8.5 & 1100 
& $1.67\times 10^{-7}$\tablenotemark{f} & 1.24 & 1400 & 110000 \\
940907 & 3159 & 18.2 & 22\tablenotemark{d} & n & 27\% & 7.0 & 2400 
& $1.42\times 10^{-6}$\tablenotemark{f} & 12.3 & 2900 &  24000 \\
950531 & 3611 &  3   & 23 & n & 71\% & 7.1 & 3600 
& $2.44\times 10^{-7}$\tablenotemark{h} & 4.43 & 4400 & 110000 \\
950907 & 3779 &  7   & 35 & n & 71\% & 7.5 & 2200 
& $3.89\times 10^{-7}$\tablenotemark{h} & 15.7 & 2700 &  27000 \\
950918 & 3805 & 40   & 20\tablenotemark{d} & n & 78\% & 7.7 & 2200 
& $8.21\times 10^{-7}$\tablenotemark{f} & 3.80 & 2800 &  67000 \\
950922 & 3814 &  5   & 46 & n & 72\% & 7.2 & 3800 
& $1.04\times 10^{-6}$\tablenotemark{h} & 4.68 & 4600 & 210000 \\
951117 & 3909 & 25   & 25\tablenotemark{d} & y & 90\% & 6.8 & 3100 
& $3.15\times 10^{-6}$\tablenotemark{f} & 29.5 & 3800 &  15000 \\
951124 & 3918 & 150  & 29\tablenotemark{d} & y & 81\% & 7.7 & 1800 
& $1.56\times 10^{-5}$\tablenotemark{f} & 50.5 & 2200 &   5800 \\
951208 & 3936 &  3.5 & 20 & y & 48\% & 6.5 & 6000 
& $2.67\times 10^{-6}$\tablenotemark{g} & 3.39 & 7400 & 200000 \\
951220 & 4048 & 17   & 15\tablenotemark{d} & y & 95\% & 7.9 & 2600 
& $1.24\times 10^{-5}$\tablenotemark{f} & 42.2 & 3200 &   5200 \\
\enddata

\tablecomments{(a): $photons~cm^{-2}~s^{-1}$ optical flux; (b):
integrated fluence up to time of first GROCSE image; (c): fit constant
for bremsstrahlung spectrum ($photons~cm^{-2}~s^{-1}$); (d): first
image taken while burst was still above BATSE threshold; (e): fluence
from 3rd BATSE Catalog; fluence calculated from spectral fit to (f):
HERB data; (g): SHERB data; (h): STTE data.}

\end{deluxetable}

\end{document}